\documentstyle[aps,twocolumn,prl,graphicx]{revtex}
\begin{document}                
\draft
\wideabs{
\author{A. I. Lvovsky\cite{Lvovsky}, S. A. Babichev}
\address{Fachbereich Physik, Universit\"at Konstanz, D-78457 Konstanz, Germany}
\date{\today}
\title{Synthesis and tomographic characterization of the displaced Fock state of light}

\maketitle

\begin{abstract}
Displaced Fock states of the electromagnetic field have been
synthesized by overlapping the pulsed optical single-photon Fock
state $|1\rangle$ with coherent states on a high-reflection
beamsplitter and completely characterized by means of quantum
homodyne tomography. The reconstruction reveals highly non-classical
properties of displaced Fock states, such as negativity of the Wigner
function and photon number oscillations.  This is the first time
complete tomographic reconstruction has been performed on a highly
non-classical optical state.
\end{abstract}
 }

\paragraph{Introduction}
Displaced Fock states (DFS) are highly non-classical states of
harmonic oscillators generated by acting upon Fock (number) states
with displacement operators. First theoretically described in 1985
\cite{Venkata}, these states have received widespread attention in
the last decade. Apart from a number of interesting and unusual
physical properties DFS possess \cite{Kral,OKKB,MC,Egypt}, it was
shown that a random quantum state's quasiprobability distribution
can be represented in terms of series of displaced number states
\cite{CG,Wuensche,MCK}. This feature has then been used
experimentally for measuring the Wigner functions of some states
of propagating electromagnetic field \cite{BW96,BW_PRA99} as well
as for characterization of motional quantum states of trapped
atoms \cite{Be}. Recently, this technique was extended to
measuring the quantum state of electromagnetic field in a
high-finesse cavity \cite{Davidovich,Haroche00}. The DFS
representation of the Wigner function has also lead to a
surprising theoretical prediction \cite{BW98} and an experimental
demonstration \cite{KWM} of quantum nonlocality of the original
Einstein-Podolsky-Rosen state.

Although DFS have proven appealing to theoreticians and useful for
various applications, their experimental characterization has not
been reported yet, the main reason being the lack of technology of
their synthesis. Recent progress in quantum reconstruction of the
single-photon Fock state \cite{Fock} has removed this obstacle,
opening up the path to generation and characterization of more
complex quantum states of light. In the present work we have
produced displaced single-photon Fock states of the light field
and investigated them by means of quantum optical homodyne
tomography \cite{HomoTomo}. As in our previous experiment
\cite{Fock}, we have observed classically impossible negative
values of the Wigner functions. In addition, we have seen photon
number oscillations that are viewed as signatures of highly
non-classical behavior typical for displaced Fock states
\cite{OKKB}.

An important difference between the undisplaced and displaced Fock
states is that the latter do carry phase information and complete,
phase-sensitive quantum tomography is therefore required for their
reconstruction. This work presents the first application of this
method to a highly non-classical state of light.

The action of the displacement operator $\hat
D(\alpha)\equiv\exp(\alpha\hat a^\dagger-\alpha^*\hat a)$ upon a
random quantum state $|\psi_0\rangle$ is best visualized in terms
of phase-space quaziprobability distributions. By acting on the
state $|\psi_0\rangle$, the displacement operator simply {\it
shifts} its Wigner function $W_{|\psi_0\rangle}(X,P)$ by the
amount of displacement:
\begin{equation}
W_{\hat D(\alpha)|\psi_0\rangle}(X,P)=W_{
|\psi_0\rangle}(X-X_0,P-P_0),
\end{equation}
where $X_0$ and $P_0$ are the real and imaginary components of the
displacement amount, $\alpha=X_0+iP_0$. For example a coherent
state is a displaced vacuum state: $|\alpha\rangle=\hat
D(\alpha)|0\rangle$.

\begin{figure}
\begin{center}
\includegraphics[width=0.3\textwidth]{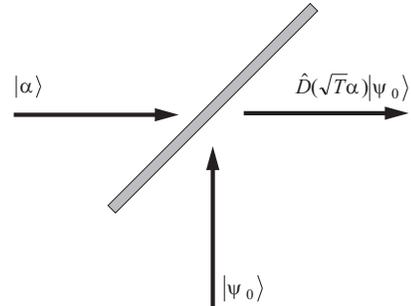}
\caption{Optical implementation of the displacement operator. The
beamsplitter must be highly reflective}
\end{center}
\end{figure}

It was shown \cite{BW96,Vogel96} that displacing a state of light
can be implemented experimentally by overlapping it with a strong
coherent state $|\alpha\rangle$ upon a highly reflecting
beamsplitter (Fig.\,1). The quantum state emerging in the
reflection (with respect to $|\psi_0\rangle$) beamsplitter output
port is to a good approximation given by
\begin{equation}
|\psi\rangle\approx\hat D(\sqrt{T}\alpha)|\psi_0\rangle,
\end{equation}
where $T\to0$ is the beamsplitter transmission.

\begin{figure}
\begin{center}
\includegraphics[width=0.45\textwidth]{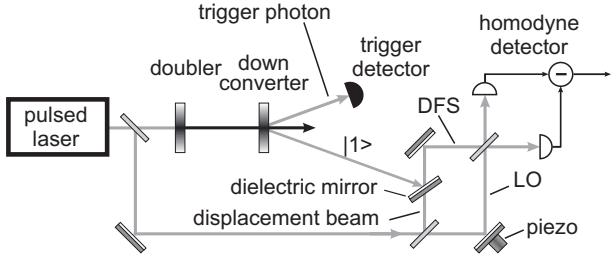}
\caption{Experimental setup. The role of a high-reflection
beamsplitter is played by a regular dielectric mirror.}
\end{center}
\end{figure}

In our work, the displacement operator was applied to the
single-photon Fock state $|1\rangle$. The optical pulses
containing single photons were prepared using conditional
measurements on the entangled photon pair generated by means of
parametric down-conversion. The two generated photons were
separated into two emission channels according to their
propagation direction (Fig.\,2). A single-photon counter was
placed into one of the channels to detect photon pair creation
events. Registration of this (trigger) photon causes the biphoton
to collapse, projecting itself onto the single-photon state in the
other down-conversion channel.

The setup for preparing the single-photon states was similar to
that described in our earlier papers \cite{Fock,Shapiro}. A 82-MHz
repetition rate train of 1.6-ps pulses generated by a
Spectra-Physics Ti:Sapphire laser at 790 nm was frequency doubled
and subsequently down-converted in a beta-barium borate crystal.
The down-conversion occurred in a type-one frequency-degenerate,
but spatially non-degenerate configuration. Pair production events
were detected via a Perkin-Elmer SPCM-AQR-14 single-photon
counting module as a rate of 300--400 pairs per second.

The prepared pulsed Fock states were then subjected to optical
displacement by means of a dielectric mirror with the reflectivity
of at least 99.99\% and a relatively strong coherent beam entering
the mirror from the rear. The electric field quadratures of the
displaced Fock states were then measured by means of balanced
homodyne detection \cite{FockHD}.

These measurements required a local oscillator which had to be
phase stable with respect to the displacement field. This was
achieved by having both fields originate from a single master
laser beam split just before the displacement mirror. Since the
two beams were recombined shortly thereafter on the 50-\%
beamsplitter of the homodyne detector, the optical arrangement
resembled a Mach-Zehnder interferometer (Fig.\,2) with a $\sim20$
cm arm length. Such compact design ensured mutual interferometric
stability of the local oscillator and the displacement field. The
interference observed between the transmitted portion of the
displacement pulse and the local oscillator allowed us to optimize
spatial and temporal mode matching between the two. Since the
optical mode of the local oscillator was also matched to the Fock
states \cite{Fock,MMpaper}, all three modes were matched to each
other as required for this measurement.

The local oscillator phase was swept by applying a linear voltage
ramp to a piezoelectric transducer upon which the local oscillator
mirror was mounted. In an experimental run that lasted a few
minutes several tens of thousands quadrature noise samples were
acquired. These data were scaled according to the vacuum state
quadrature noise measured at the same local oscillator intensity.

The scaled experimental data, along with the reconstructed Wigner
functions, are shown in Fig.\,3. Fig.\,3(a) represents the vacuum
state $|0\rangle$ acquired with both the single-photon and the
displacement beam blocked. Invoking the displacement operator
$\hat D(\alpha)$ with $\alpha=0.60$ converts the vacuum state into
a coherent state $|\alpha\rangle$ [Fig.\,3(b)]. Fig.\,3(c)
corresponds to the ensemble $\hat\rho_{\rm
mix}=\eta|1\rangle\langle 1|+(1-\eta)|0\rangle\langle 0|$ acquired
with the single-photon channel unblocked. The admixture of the
vacuum occurs due to a number of setup imperfections \cite{Fock};
the quantum efficiency of this measurement was equal to
$\eta=0.62$. The displacement operator transformed this ensemble
into a mixture of the DFS and a coherent state of the same
amplitude [Fig. 3(d)],
\begin{equation}
\hat\rho_{\rm disp}=\eta\hat D(\alpha)|1\rangle\langle 1|\hat
D^\dagger(\alpha)+(1-\eta)|\alpha\rangle\langle\alpha|.\label{rho_disp}
\end{equation}
The Wigner function of this ensemble is negative around the point
$X+iP=\alpha$ as long as $\eta>0.5$.

\begin{figure}
\begin{center}
\includegraphics[width=0.45\textwidth]{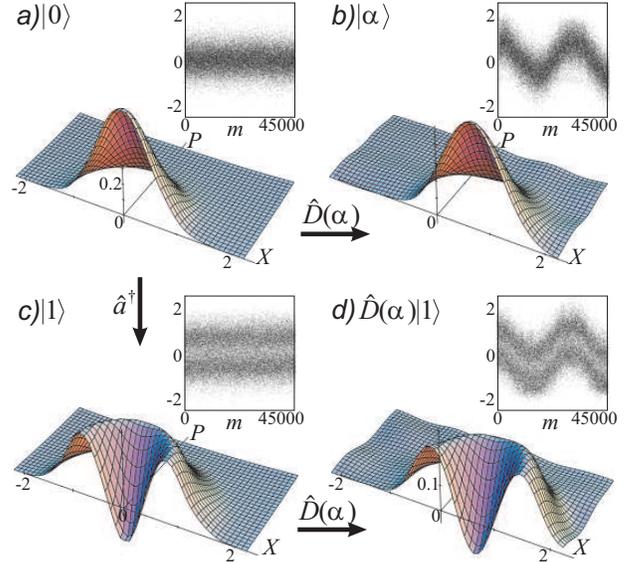}
\caption{Raw (normalized) homodyne output data and reconstructed
Wigner functions. (a) vacuum state; (b) coherent state; (c) Fock
state; (d) displaced Fock state. In (b) and (d), $\alpha=0.60$ }
\end{center}
\end{figure}

The phase-averaged Wigner functions shown in Fig.\,3(a,c) were
calculated by means of Abel transformation. Reconstruction of the
phase-sensitive Wigner functions [Fig.\,3(b,d)] was performed via
the inverse Radon transformation. To this end, each individual
quadrature noise sample $X_m$ was associated with a particular
local oscillator phase value $\theta_m$, determined by analyzing
the periodic behavior of the raw data pattern; due to a large
number of acquired data the dependence of $\theta_m$ on $m$ could
be assumed linear. We have then applied the standard filtered
back-projection algorithm \cite{Leonhardt} $directly$ to the set
of normalized quadrature noise samples in terms of an empirical
average:
\begin{equation}
W(X,P)=\frac{1}{\pi}\langle {\rm
K}(X\cos\theta_m+P\sin\theta-X_m)\rangle,
\end{equation}
where ${\rm K}(x)=\cos(k_cx)+k_cx\sin(k_cx)-1)/x^2$ is the
transformation kernel, $k_c=6.4$ is the cutoff frequency and
$\theta_m$ ranges between 0 and $2\pi$. This approach is simpler
and more precise than the traditional one involving an
intermediate step of binning up the data and calculating
individual marginal distributions associated with each phase.

\begin{figure}
\begin{center}
\includegraphics[width=0.35\textwidth]{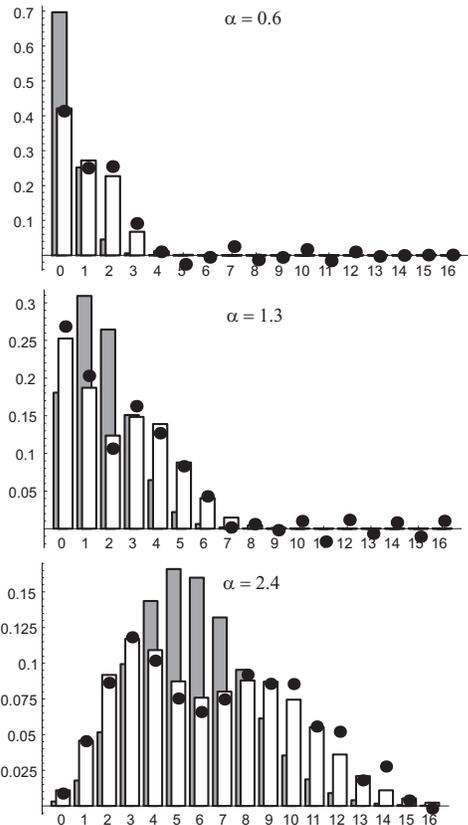}
\caption{Diagonal elements of the displaced Fock state density
matrices in the Fock representation. Black dots: experimental
values determined via the quantum state sampling technique. White
columns: theory for given $\alpha$ and $\eta$. Grey columns:
theory for coherent states with the same $\alpha$. }
\end{center}
\end{figure}

We have also evaluated the diagonal elements of the DFS density
matrices in the Fock representation by applying the standard
quantum state sampling method
\cite{Leonhardt,patterns1,patterns2}. As displayed in Fig.\,4, the
distributions associated with high amplitudes exhibit two peaks.
This is a well-known characteristic feature of displaced Fock
states, with the number of peaks for the state $\hat
D(\alpha)|n\rangle$ being equal to $n+1$ \cite{OKKB}. The
theoretical fits shown were obtained by substituting the photon
statistics of the displaced Fock and coherent states into the
ensemble (\ref{rho_disp}), with the values of $\alpha$ and $\eta$
determined by analyzing the raw data. Note that the classical
noise generated by the homodyne detector at large displacements
has lowered the effective quantum efficiency from the values
observed in undisplaced Fock states in the same experimental run.
The amount of reduction was equal to 0.02 and 0.10 for
$\alpha=1.3$ and $\alpha=2.4$, respectively,

In conclusion, we have synthesized a new quantum state of light, the
displaced Fock state, and reconstructed its Wigner function and
density matrix by means of optical homodyne tomography. This state
exhibits highly nonclassical features such as negativity of its
Wigner function and photon number oscillations. A novel approach to
the inverse Radon transformation was employed.

In the present work we investigated the quantum state in only one
of the beamsplitter output channels, discarding the other one. In
a future publication \cite{cat} we subject both channels to
quantum measurements and study their nonclassical correlations
demonstrating the entangled nature of the two-mode state emerging
from the beamsplitter.

A. L. was supported by the Alexander von Humboldt Foundation.
Support for this work was provided by the Deutsche
Forschungsgemeinschaft and Optik-Zentrum Konstanz.

\end{document}